# Light emission, light detection and strain sensing with nanocrystalline graphene

Adnan Riaz[1,2], Feliks Pyatkov[1,3], Asiful Alam[1,3], Simone Dehm[1], Alexandre Felten[4], Venkata S. K. Chakravadhanula[1,6], Benjamin S. Flavel[1], Christian Kübel[1,6,7], Uli Lemmer[2,5] and Ralph Krupke[1,3]

[1]Institute of Nanotechnology, Karlsruhe Institute of Technology, 76021 Karlsruhe, Germany
[2]Light Technology Institute, Karlsruhe Institute of Technology, 76021 Karlsruhe, Germany
[3]Department of Materials and Earth Sciences, Technische Universität Darmstadt, 64287 Darmstadt, Germany
[4]Research Center in Physics of Matter and Radiation, University of Namur, Namur, Belgium
[5]Institute of Microstructure Technology, Karlsruhe Institute of Technology, 76021 Karlsruhe, Germany
[6]Helmholtz Institute Ulm, 89081 Ulm, Germany
[7]Karlsruhe Nano Micro Facility, Karlsruhe Institute of Technology, 76021 Karlsruhe, Germany
E-mail: krupke@kit.edu

## Abstract

Graphene is of increasing interest for optoelectronic applications exploiting light detection, light emission and light modulation. Intrinsically light-matter interaction in graphene is of a broadband type. However by integrating graphene into optical micro-cavities also narrow-band light emitters and detectors have been demonstrated. The devices benefit from the transparency, conductivity and processability of the atomically thin material. To this end we explore in this work the feasibility of replacing graphene by nanocrystalline graphene, a material which can be grown on dielectric surfaces without catalyst by graphitization of polymeric films. We have studied the formation of nanocrystalline graphene on various substrates and under different graphitization conditions. The samples were characterized by resistance, optical transmission, Raman, X-ray photoelectron spectroscopy, atomic force microscopy and electron microscopy measurements. The conducting and transparent wafer-scale material with nanometer grain size was also patterned and integrated into devices for studying light-matter interaction. The measurements show that nanocrystalline graphene can be exploited as an incandescent emitter and bolometric detector similar to crystalline graphene. Moreover the material exhibits piezoresistive behavior which makes nanocrystalline graphene interesting for transparent strain sensors.

## 1. Introduction

Graphene has been extensively explored over the last decade and many aspects of its unique properties have been revealed and investigated [1]. Besides being a promising transparent, conductive and flexible coating material it is the material's potential for electronics and optoelectronics which attract a lot of attention [2]. In particular the light-matter interaction in graphene is of interest from a fundamental but also application perspective [3]. It has been shown that graphene can be operated as a broad-band light detector or emitter and the mechanism of photocurrent generation and light emission has been revealed as bolometric or electrothermal and thermal, respectively [4,5,6,7]. More recently narrow-band light emission and detection has been obtained by photonic engineering through integration into an optical microcavity [8,9]. These devices were based on exfoliated or CVD-grown graphene with domain sizes on the order of micrometers. Thereby manual transfer processing was involved which still poses a challenge for reliable, wrinkle-free wafer-scale device fabrication [10]. Considering the light-matter interactions involved it is however not clear whether graphene with large crystalline domains would be required for operation. Recently the growth of thin carbon films on dielectric surfaces by graphitization of polymeric films has been demonstrated [11,12]. The simple, metal-catalyst free process yields graphene with nanometer-scale domain size, often termed nanocrystalline graphene [13]. Although the low-bias electrical transport of nanocrystalline graphene is distinctly different from crystalline graphene, the material is conducting and transparent and should be suitable for some of the potential optoelectronic applications [14]. Recently nanocrystalline graphene has been used as transparent electrode in a silicon photodiode and an organic solar cell [15,12]. Here in this work we synthesize, characterize and integrate nanocrystalline graphene into devices, and investigate its feasibility for nanoscale light emission, nanoscale light sensing, and in addition for strain sensing. We have studied first the formation of nanocrystalline graphene on various substrates and graphitization conditions, and characterized the material by resistance, optical transmission, atomic force microscopy (AFM), X-ray photoelectron spectroscopy (XPS), Raman and transmission electron microscopy (TEM) measurements. The conducting and transparent wafer-scale material with nanometer grain size was then patterned and integrated into devices for studying light-matter interaction and strain sensing.

## 2. Experimental details

Figure 1 shows schematically the process flow. For the synthesis of nanocrystalline graphene (NCG) we have modified the approach of Zhang and co-worker [11] by spin-casting photoresist onto various substrates and graphitizing the material under high vacuum at high temperatures. As substrates we have used p-doped-Si (<100>, Bor, $\rho < 0.005$ Ωcm, thickness 0.381-0.525 mm) with 100 nm, 300 nm and 800 nm thermal oxide from Active Business Company GmbH; n-doped-Si (<100>, P, $\rho > 1000$ Ωcm, thickness 0.525 mm) with 500 nm $Si_3N_4$ on top of 800 nm thermal oxide from Active Business Company GmbH; Quartz (<0001>, z-cut ±0.3°, thickness 0.5 mm) from Alineason Materials & Technology GmbH; and HOQ 310 Quartzglas (thickness 0.5 mm) from Aachener Quarz-Glas Technologie Heinrich GmbH. The lateral dimensions of the substrates were 1x1 $cm^2$. The substrates were cleaned for ~10 s in a sonication bath with Acetone, rinsed with Isopropanol (IPA) and dried in a Nitrogen stream. Subsequently the substrates were exposed to an oxygen plasma (1 min, 0.2 mbar, 25 sccm $O_2$, 30 % power) using an Atto LC PC from Diener electronic GmbH. The clean substrates were then pre-baked for 2 min on a hot plate at 110° C. After cool-down 30 µl of photoresist solution was spin-casted onto the substrate under ambient conditions using 8000 rpm for 30 s. The substrates were then soft-baked for 1 min at 110° C. The photoresist solution was prepared by diluting Microposit S1805 positive photoresist from Rohm&Haas with Propylene Glycol Monomethyl Ether Acetate (PGMEA) of 99 % purity from Sigma Aldrich GmbH. PGMEA is the main component of S1805. We have used S1805: PGMEA dilution ratios of 1:14, 1:16, 1:18 and 1:21 to target NCG thicknesses of 6 nm, 4 nm, 2 nm and 1 nm, respectively. The polymer films were graphitized by loading them into a Gero SR-A 70-500/11 high-temperature oven equipped with a Quartz glass tube, a turbo-pump vacuum system and a temperature controller unit. After evacuation to ≤ $10^{-6}$ mbar the temperature was raised to 100 °C for ~ 30 min to promote water desorption. The temperature was then raised to 1000 °C at a rate of 10°C/min and stabilized for 10 h. The samples were cooled-down by switching-off the heater and released to the ambient at temperatures below 100° C.

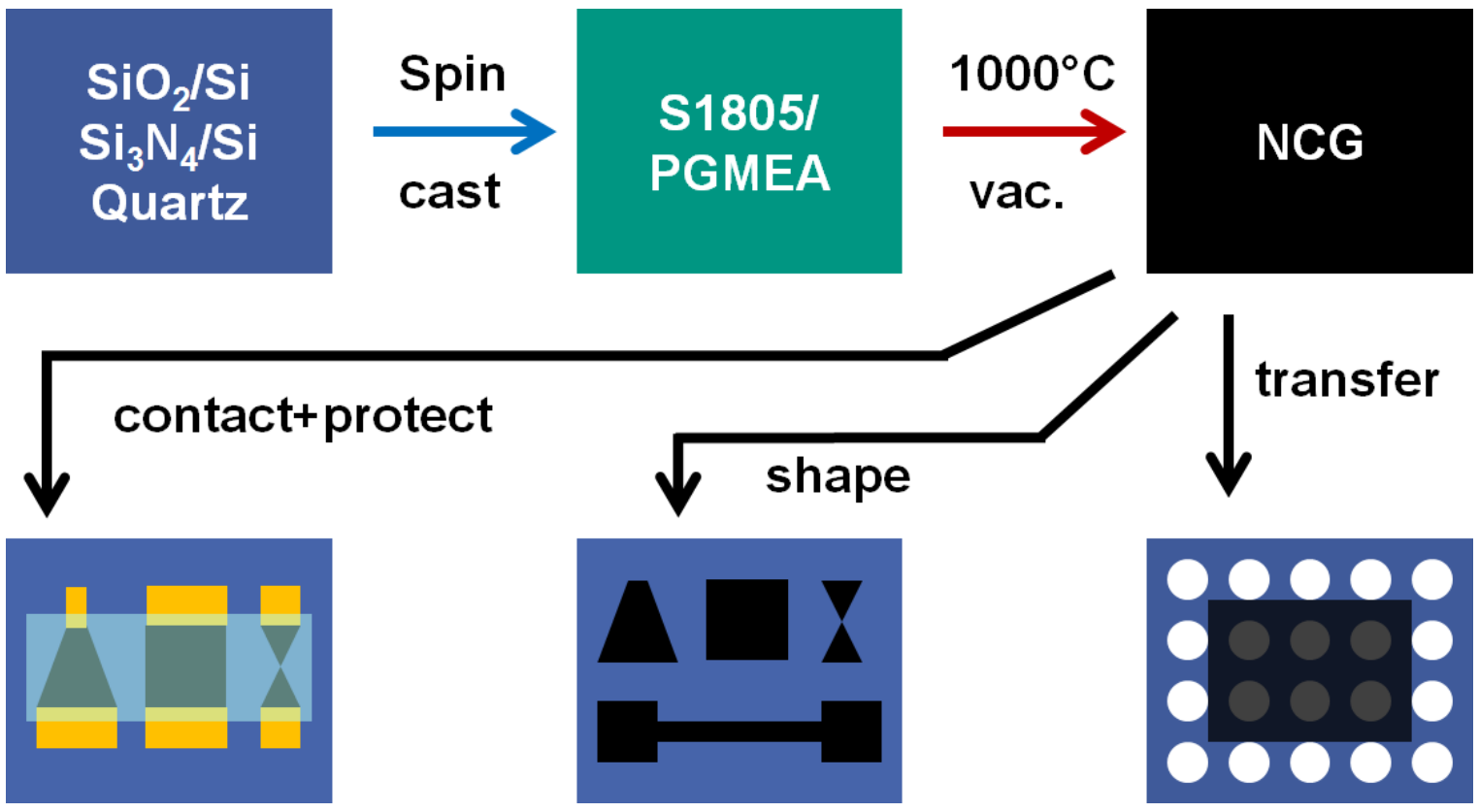

**Figure 1.** Schematic process flow. Substrates were spin-casted with a resist-solvent mixture and graphitized at high-temperature under high-vacuum, as described in the text. The wafer-scale nanocrystalline graphene (NCG) was characterized by Raman spectroscopy, X-ray photoelectron spectroscopy, sheet-resistance measurements and optical transmission spectroscopy. For transmission electron microscopy and diffraction the NCG was transferred onto a grid. Using lithography methods the NCG was patterned for atomic force microscopy and piezoresistivity measurements. Photocurrent and light emission measurements were performed on NCG with electric contacts and a dielectric coating.

Before further processing, the NCG samples were characterized by measuring the sheet resistance with the van-der-Pauw method [16] in a Cascade probe station using gold-plated needles and an Agilent 4155C semiconductor parameter analyzer. Raman spectra were measured with a Witec CRM 200 confocal microsope under 1mW laser excitation at 632.8 nm with a Nikon Plan 100x objective (NA = 0.9). X-ray photoelectron spectra (XPS) were measured with an Escalab 250Xi system from Thermo. XPS spectra of the NCG films were acquired with a monochromatic AlKα source (1486.6 eV), using a spot size of 400 μm and a pass energy of 20 eV. A flood gun was used to prevent charging of the dielectric substrates. Transmission spectra of NCG on quartz were measured in a Cary 500 UV-Vis-NIR spectrometer.

For further characterization the NCG samples had to be processed as indicated in Figure 1. For transmission electron microscopy (TEM) characterization a 2 nm thin NCG layer synthesized on 800 nm-$SiO_2$/Si was transferred onto a Quantifoil holey carbon grid using a direct transfer process [17]. The TEM grid was placed on the sample and a few drops of IPA were added. Due to the hydrophobic nature of both the TEM carbon grid and the NCG, the grid adhered to the NCG layer. The thermal oxide layer was then etched by 2 % hydrofluoric acid (HF) overnight. After etching, the TEM grid was rinsed with deionized water (DI) and with IPA. A Zeiss Ultra Plus scanning electron microscope (SEM) was used to confirm the transfer process and to image samples throughout subsequent processing steps. The images were recorded with the in-lens detector at 5 keV beam energy. HRTEM images and selected area electron diffraction (SAED) patterns were acquired using an aberration (image) corrected FEI Titan 80-300 operated both at 80 kV and 300 kV acceleration voltage. EELS spectra were recorded at 80kV. The SAED patterns were processed using the Digital Micrograph (Gatan) script PASAD [18].

For characterizations that required laterally patterned NCG, the samples were locally etched by oxygen plasma. The areas of interest were defined by electron beam (ebeam) lithography and were protected by an aluminum layer. First, Tungsten markers were defined by spin-casting 950k PMMA ebeam resist from Allresist at 5000 rpm for 60 s and baking at 160 °C for 30 min, thus forming a 200 nm thin PMMA layer. The PMMA was exposed by 30 keV electrons using a Raith Elphy Plus pattern processor system attached to a LEO1530 SEM. The exposed PMMA was developed in Methyl Isobutyl Ketone (MIBK) and IPA at a ratio of 1:3 for 35 s. The markers were then formed by sputtering 40 nm Tungsten in a Bestec custom design sputtering system (400W DC, 1 min), subsequently lifted-off in Acetone and rinsed with IPA. The NCG was then synthesized by the procedure outlined before. The NCG was then laterally structured with respect to the position of the Tungsten markers in a second lithography step by forming, exposing and developing PMMA under conditions identical to the first lithography step. 15 nm Al was evaporated as an etch resist. After lift-off in Acetone and IPA, the unprotected NCG was etched in an oxygen plasma (3.5 min, 60 mtorr, 20 sccm $O_2$, 30 W) using an Oxford Plasmalab 80 Plus reactive ion etching system (RIE). Subsequently the Al resist layer was etched in 3% metal ion free tetramethylammonium hydroxide (MIF 726) for 30 s. The patterned NCG was then characterized with a Bruker dimension icon atomic force microscope (AFM) in tapping mode to determine the NCG thickness and surface roughness. Piezoresistance measurements of NCG on 800nm-$SiO_2$/Si were taken in a custom designed three-point bending fixture with laser-assisted zero-point adjustment. The sample was electrically contacted by fixing flexible Au wires (Ø = 0.2 mm) using Indium disks (Ø = 2 mm). The strain induced by extending the central cylindrical post was calculated based on the distance of the cylindrical fixed posts (17.2 mm) and the substrate thickness (0.381 mm), translating into 0.1 % strain per 200 µm vertical movement of the central post. Uniform curvature has been assumed.

For photocurrent and incandescence experiments the NCG was electrically contacted and subsequently coated by a dielectric using a third and a fourth lithography step, respectively, by forming, exposing and developing PMMA under conditions identical to the previous lithography steps. The electrical contacts to NCG were made by sputtering 70 nm Tungsten and lifting-off in Acetone and IPA. A protecting layer of 30 nm $Al_2O_3$ was formed in a Savannah 100 atomic layer deposition system (ALD) from Cambridge NanoTech, using 273 cycles of alternating exposures to 0.02 pulse/sec Trimethyl Aluminum (TMA) and 0.02 pulse/sec water vapor at 200 °C. In a final lithography step using PMMA as etch resist the $Al_2O_3$ on top of the Tungsten contact pads was etched within the RIE (6.5 min, 15 mtorr, 40 sccm Ar + 10 sccm $CHF_3$, 200 W), to ensure good electric contact between probe needles and the Tungsten pads. Photocurrent maps have been measured using a NKT SuperK Extreme EXW-6 supercontinuum light source coupled through a NKT SuperK Select acousto-optic tunable filter, a NKT FD-9 fiber, and a reflective beam collimator to a Zeiss Axiotech Vario microscope, equipped with a beam-splitter to guide 90% of the light through a Mitutoyo Plan Apo NIR 100x/0.50 objective onto the sample. The remaining 10% intensity is used for in-situ monitoring of intensity fluctuations during measurement. Samples were mounted on a Standa 8MTF-102LS05 motorized stage (stepsize 375 nm) and contacted with probe needles. The backreflected light was measured with a Silicon Sensor PC-50-6 silicon PIN photodiode, to correlate the photocurrent signal with the sample topography. The photocurrent signal was detected using a SR 830 lock-in amplifier and a SR 570 pre-amplifier. The AOTF was used to select the excitation wavelength and the chopping frequency to 1065 nm and 2.3 kHz at 600 μW irradiation power. For light emission measurements, a Zeiss AxioTech Vario microscope, directly attached to an Acton Research SpectraPro 2150i spectrometer and a Princeton Instruments PIXIS 256E Silicon CCD camera (1024 × 256 pixels, −60°C) with in a light-tight box was used. The spectrometer can operate in the imaging mode, with a mirror to take real-space images, or in the spectroscopy mode, with a diffraction grating (300 grooves/mm, 750 nm blaze wavelength). The samples were mounted on a Standa 8MTF-102LS05 motorized stage, electrically contacted with probe needles and biased with a Keithley 6430 SourceMeter. Incandescence images and images under external illumination were recorded with a Mitutoyo Plan Apo 80×/0.50 objective.

## 3. Results and discussion

Figure 2a shows an AFM image across an edge of a structured 1 nm thin NCG layer on 800nm-$SiO_2$/Si. The data shows that the surface roughness of NCG is similar to the roughness of the $SiO_2$ surface, indicating a conformal coating of the substrate. The edge appears straight and shows no sign of underetching. The NCG thickness was adjusted by diluting the resist to yield a targeted nominal thickness. The degree of control is shown in figure 2b. The sheet resistance of the NCG is typically on the order of 20-80 kΩ/sq., depending on the NCG thickness and the type of substrate, as can be seen in figure 2c.

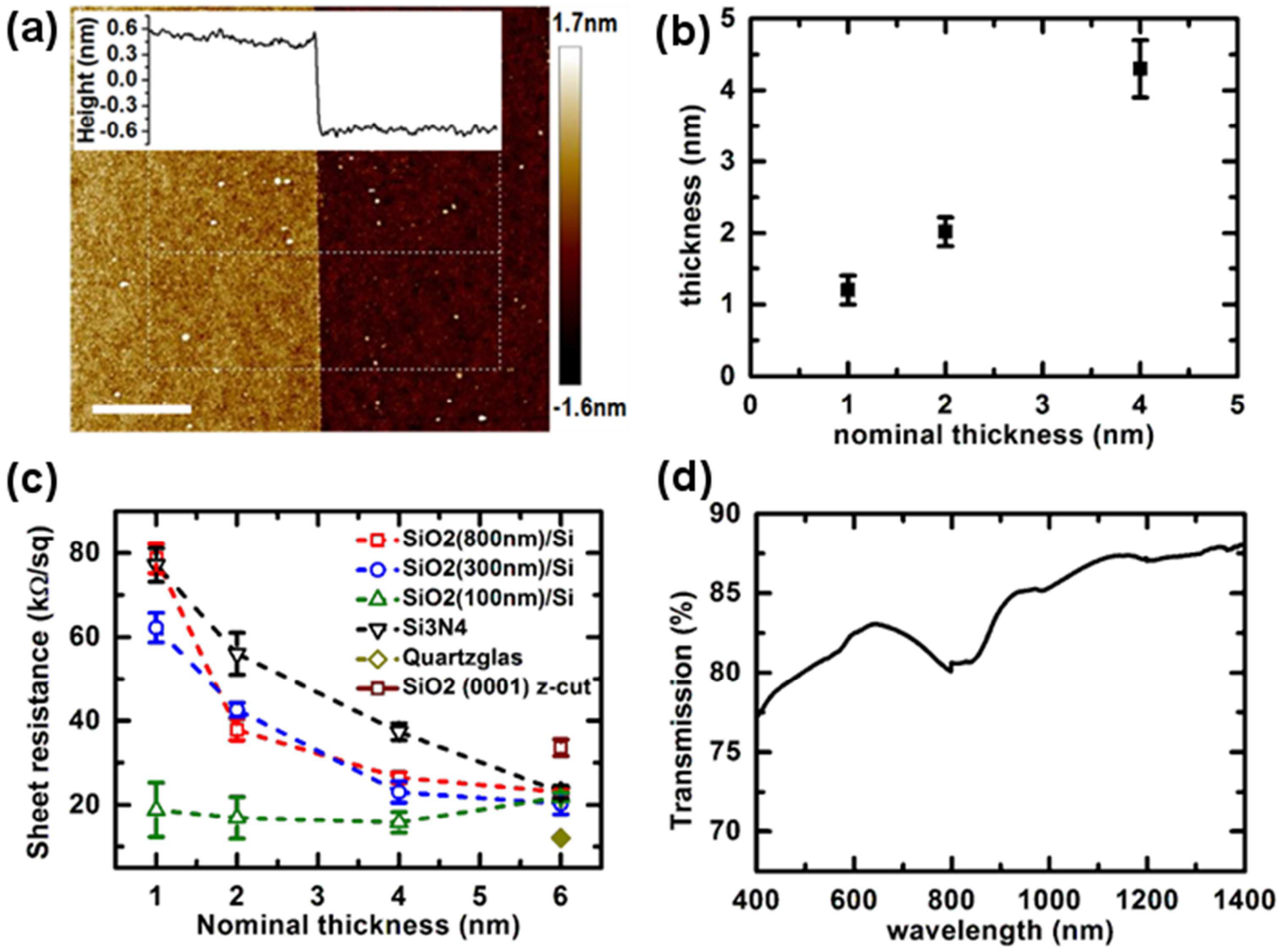

**Figure 2**: (a) Atomic force microscopy image of a patterned 1 nm thin NCG layer on 800nm-$SiO_2$/Si substrate. The inset shows an averaged cross-section of the indicate area. Scale bar equals 1 μm. (b) Measured NCG thickness versus its nominal thickness. (c) Sheet resistance versus nominal thickness of NCG on various substrates as indicated. (d) Transmission spectrum of 6 nm thick NCG on <0001> z-cut Quartz. All NCG samples have been synthesized at 1000°C@10h.

The mean value and the error bars were obtained from measurements on 3-5 samples for each data point. Overall the data shows the expected dependence of the sheet resistance on the inverse layer thickness. The sheet resistance values are comparable to the 30 kΩ/sq. reported by Zhang et al. [11] for 1 nm NCG, also formed at 1000 °C albeit under reducing atmosphere. Compared to CVD-graphene [19] and carbon nanosheets [14] is the sheet resistance of our NCG 2 orders of magnitude larger and 2 orders of magnitude smaller, respectively. Obviously the carrier transport across grain boundaries has a large influence on the overall resistance. We point out that the graphitization of the resist seems to depend on the surface structure of the $SiO_2$ as can be seen by the sensitivity of the sheet resistance to the substrate. Hence the catalytic activity appears to be slightly different for Quartzglas, <0001> z-cut Quartz and 100nm-$SiO_2$ as compared to the thicker thermal oxides on Si. The optical transparency of a 6 nm thick NCG film on <0001> z-cut Quartz is shown in figure 2d. The transmission increases from 77% at 400 nm to 87 % at 1400 nm and shows an enhanced absorption around 800 nm. Normalized by the thickness, the optical transmission of NCG is comparable to that of graphene. We speculate whether the enhanced absorption at 800 nm is correlated with the crystallite size and we will come back to this point later in the discussion. To determine the composition and the hybridization of the NCG material we have analyzed the C 1s peak of the XPS signal and compared the spectrum to data recorded on Bilayer graphene (mechanically exfoliated on 90nm-SiO2/Si) and graphite. Figure 3a shows that the C 1s signal has a main peak at 284.4 eV similar to graphite, and hence assigned it to the sp-2 hybridized carbon atoms. Also the width of ~1.2 eV fits very well to Bilayer graphene which has a similar thickness as NCG.

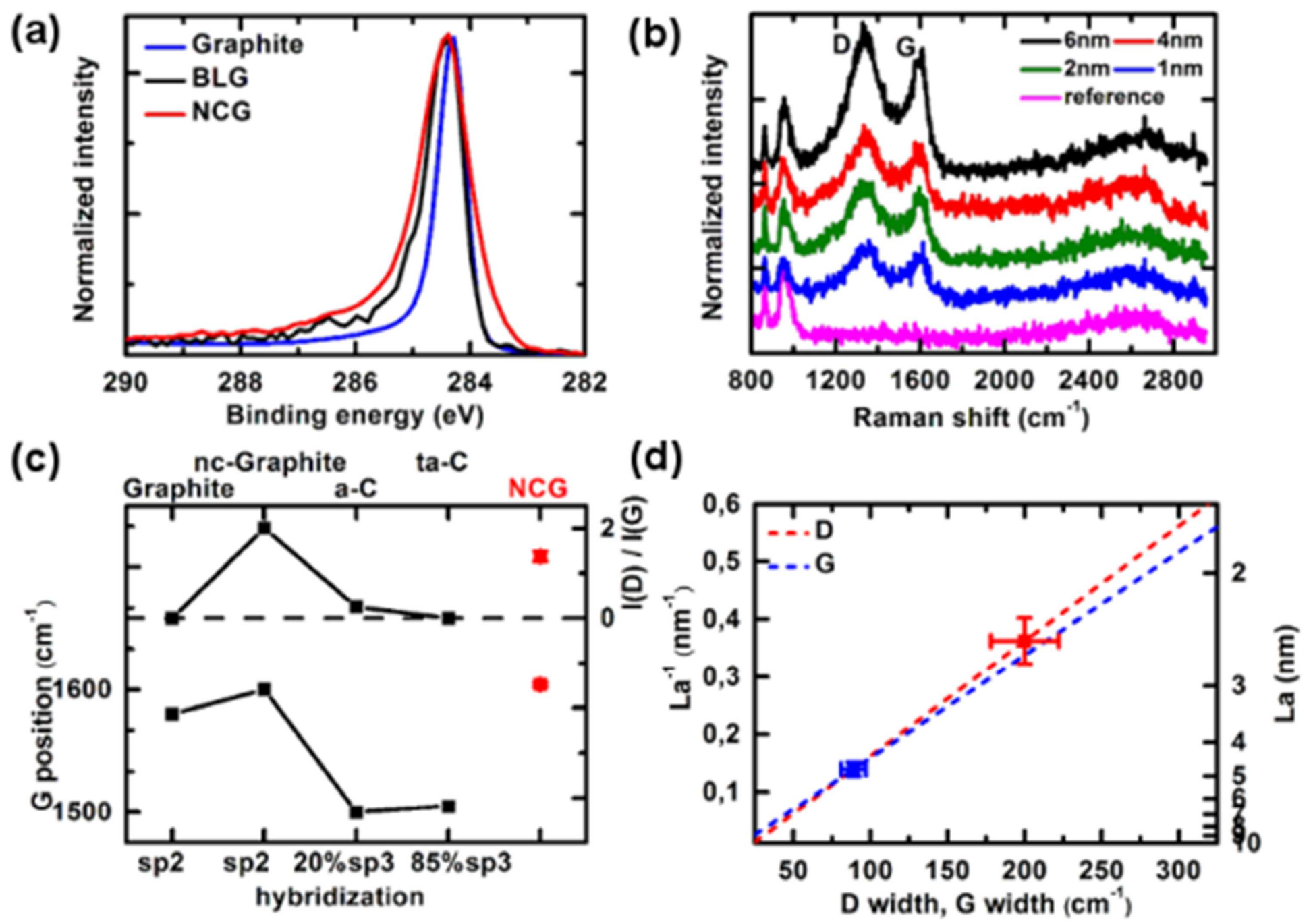

**Figure 3**: (a) XPS of 1 nm thick NCG layer on 800nm-$SiO_2$/Si, compared with bilayer graphene (BLG) and graphite. (b) Raman spectra of NCG with various thicknesses, grown on 800nm-$SiO_2$/Si. The spectra are normalized to the D peak and vertically shifted for clarity. (c) Raman G-peak position and intensity ratio of D-peak and G-peak for NCG compared with data from ref 23 for graphite, nanocrystalline graphite and two forms of DLC as explained in the text. The sp-hybridization is indicated. (d) Crystallite size *La* in NCG determined from the full-width-at-half-maximum of the D- and G-peaks. The dashed lines are extrapolated correlations based on ref. 24. The graphs (c) and (d) include a statistical evaluation of data from all samples synthesized at 1000°C.

Some peak broadening towards higher and lower binding energy is observed for NCG, which could have its origin in the nanocrystallinity of the material, similar to observations in low-crystalline carbon nanotubes and defective graphite [20,21]. In the survey spectrum (not shown), no additional elements besides Si and O from the substrate were detected. Hence we can safely conclude that our NCG is a graphitic material with a high degree of sp2 hybridization. XPS also allows determining the thickness of the carbon layer from the attenuation of the photoelectrons emanating from the $SiO_2$ [22]. We have used the Si 2p with an attenuation length of 3.5 nm [14] and obtained 0.95 nm for a 1 nm thick layer. Hence the thickness determined by XPS is consistent with the AFM measurement. All samples were characterized by Raman spectroscopy to obtain an additional confirmation on the hybridization and to determine the crystallite size *L*a in the NCG layer. Figure 3b shows Raman spectra of NCG on 800nm-$SiO_2$/Si for different nominal layer thickness. For other process conditions we refer to Fig. S1-3. Characteristic to all samples are broad D and G modes, and the absence of a clear 2D peak, similar to refs 14 and 11. We used the pioneering work of Ferrari and Robertson to determine the hybridization and crystallinity of our NCG [23]. In Figure 3c we have compared the G-peak position and the intensity ratio *I*(D)/*I*(G) of the D-peak to the G-peak of NCG with the data measured on graphite, nanocrystalline graphite, diamond-like carbon (DLC) with 20% $sp^3$-content (a-C) and DLC with 85% $sp^3$ (ta-C). The NCG data fits well to the nanocrystalline graphite with 100% $sp^2$ content, and hence confirms nicely the XPS data. From the width of the D and G modes we estimated the size of *L*a by referring to the work of Cançado et al. [24]. The D mode width corresponds accordingly to *L*a ≈ 4-5 nm, whereas the G mode width indicates *L*a ≈ 2-3 nm. To discern the difference we transferred a 2 nm thick NCG layer formed on 800nm-$SiO_2$/Si onto a TEM grid. The mean value and the spread of the ratio *I*(D)/*I*(G), the G-peak position and the width of the D and G modes have been evaluated from samples prepared on various substrates and of different NCG thicknesses (see table T1 in supporting information). Figure 4a-b are low and high resolution TEM images which show that the NCG film is continuous. Since the morphology appears to be not consistent with the AFM characterization we assume that the inhomogeneity, visible in figure 4b, is due to residues from the HF etching of $SiO_2$. An SAED image was recorded on a flat region showing continuous diffraction rings (figure 4c). We were measuring an SAED pattern of a polycrystalline material [25]. The position of the two rings at 0.12 $nm^{-1}$ and 0.21 $nm^{-1}$ match well to the (110) and (100) reflexes of in-plane oriented graphene, respectively. Although we do not see clear domains in the HRTEM image in figure 4b, the FFT of the image shows nevertheless a lattice plane distance of 0.21 $nm^{-1}$ (inset fig. 4b). A radially averaged line profile of the SAED image is shown in figure 4d, which we have analyzed to determine the average domain size. After background correction we have applied the Scherrer equation to the peak of the (100) reflex with a FWHM of 0.656 $nm^{-1}$ using a shape factor of 0.9 [26]. As a result we obtain an average domain size of 3 nm, which is closer to 2.2 nm derived for this film from the width of the Raman D mode peak compared to the G mode width. It should be noted that similar domain sizes of 2-5 nm have been reported by Turchanin et al. for organic monolayers graphitized on Au [13]. We have also measured the EELS spectrum at the carbon edge of nanocrystalline graphene (Figur S6) and observe transitions which fit to the π* and σ* states of a graphitic material [27,28].

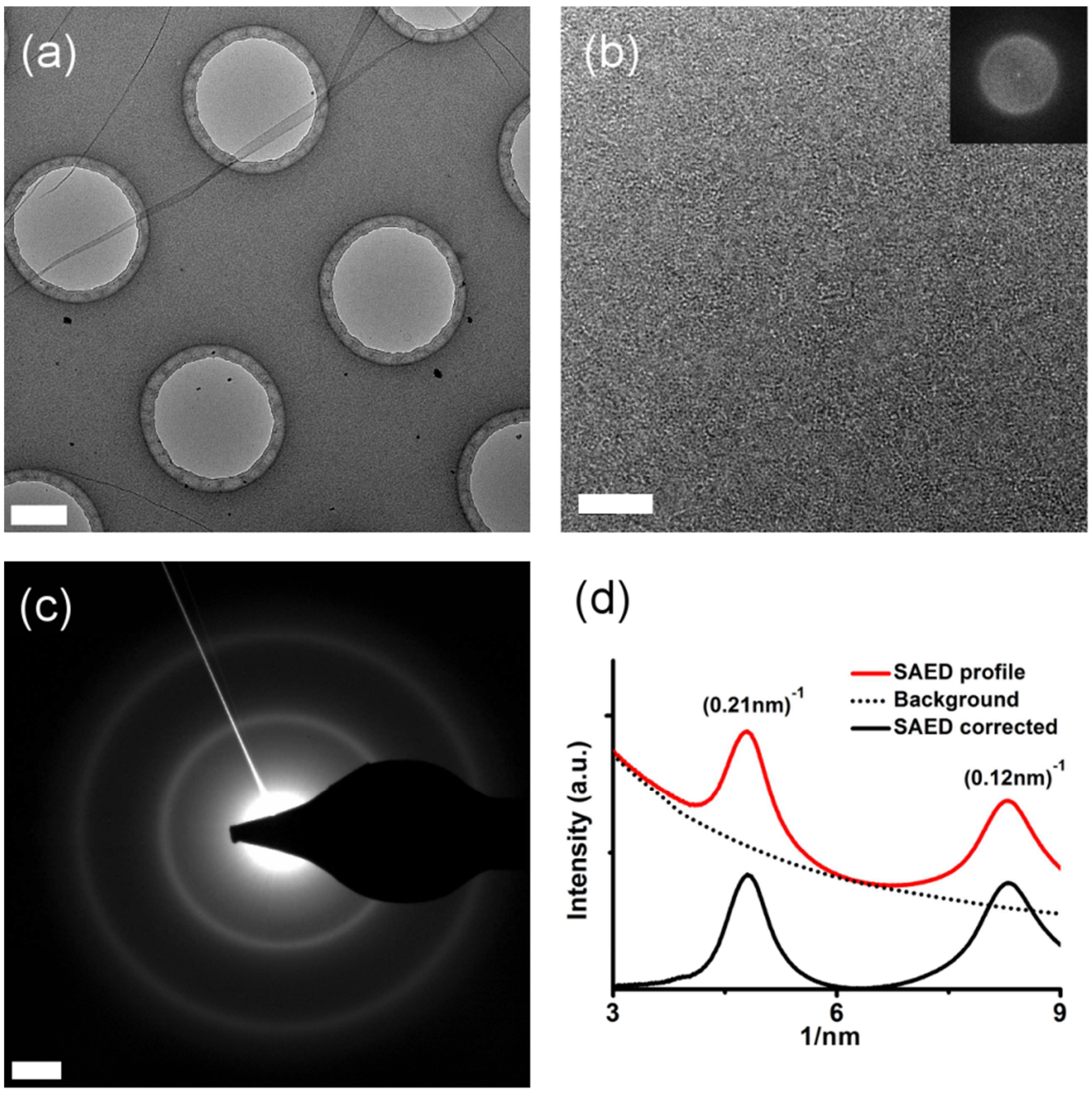

**Figure 4**: Transmission electron microscopy characterization of a 2 nm NCG layer synthesized on 800nm-SiO2/Si in 10h@1000°C. (a) Low resolution TEM image of the NCG after transfer onto a holey carbon grid. A fault can be seen in the upper left. Scalebar equals 0.4 µm. (b) High resolution TEM image measured through a hole. Scalebar equals 4 nm. (c) SAED image. Scalebar equals 2 $nm^{-1}$. (d) Radially averaged line profile of the SAED image.

We will now describe the results of the first spatially resolved photocurrent measurements on NCG and discuss the mechanism of photocurrent generation. Figure 5a shows a typical photocurrent map of a 20µm x 20µm square-shaped NCG in contact with Tungsten electrodes (overlap 1µm), recorded at +1 V source-drain voltage. We have measured simultaneously the local reflectivity (not shown) to correlate the photocurrent signal with the sample structure. Instead of superimposing the photocurrent map with the reflectivity map we have superimposed the corresponding SEM image for better visibility of the NCG and adjusted its relative position to the photocurrent map by matching to the reflectance map (see also Fig. S4). From the data we can see that the photocurrent is generated at the two contacts between NCG and Tungsten with similar magnitude. The sign of the photocurrent is phase-shifted by 180° and its magnitude increases with the applied bias. The behavior of the photocurrent is symmetric for both bias polarities as can be seen in the vertical cross-section shown in figure 5b. Figure 5c shows a photocurrent map of a 20 µm x 2 µm / 20 µm triangular-shaped NCG in contact with W electrodes (overlap 1 µm), also recorded at +1 V source-drain bias. In this case the photocurrent is primarily generated at the contact with the smaller area, where the apex of the NCG triangle is in contact with Tungsten. Again the photocurrent is phase-shifted by 180°, the magnitude increases with the applied bias and the photocurrent is symmetric with respect to the bias polarity (see figure 5d).

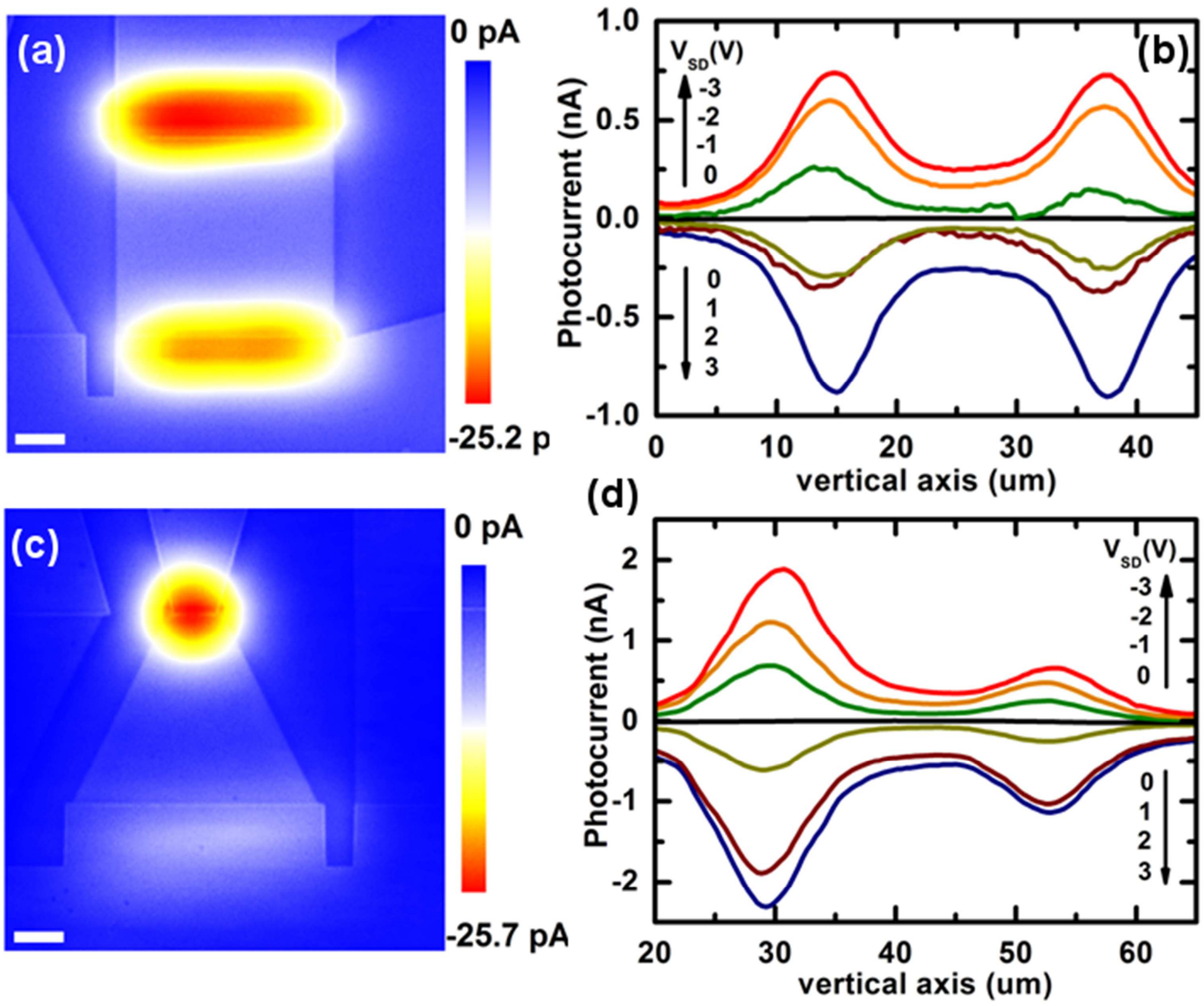

**Figure 5**: Photocurrent maps of square-shaped (a) and triangle-shaped (c) NCG devices. SEM images were superimposed as explained in the text. $V_{SD}$ = +1 V. Scale bars equal 5 µm. (b),(d) Vertical cross-section through the corresponding photocurrent maps as a function of voltage bias. The devices were fabricated from 4 nm thin NCG layers synthesized on 800nm-$SiO_2$/Si at 10h@1000°C.

The discussion on the mechanism of photocurrent generation is rather straightforward. The fact that we do not observe a photocurrent signal at zero bias excludes a significant photovoltaic or electrothermal contribution and can only be explained with a bolometric effect [29]. The temperature dependence of the NCG resistance has then to be positive $dR/dT > 0$ since the photocurrent is phase-shifted by 180° with respect to the applied bias voltage: under illumination the resistance of the sample increases and hence the total current decreases. The photocurrent becomes negative since it is the difference between the total current and the dark current. A bolometric photocurrent must also be correlated with regions of high resistance. Apparently the contact resistance between NCG and W in figure 5 is dominating the total device resistance and hence these regions show up in the photocurrent maps. For asymmetric contacts as in figure 5c it is only one contact that dominates the entire device resistance. Hence bolometric photocurrent mapping can be considered as a tool for mapping the internal device resistance. To test our hypothesis we have patterned within two opposing NCG triangles with 20µm x 10µm overall dimensions and a 0.5 µm wide and 1 µm long NCG constriction (see also Fig. S5). As in the previous devices NCG has been contacted with W (overlap 1 µm). Figure 6a shows that the photocurrent signal is now the highest at the position of the constriction and hence nicely confirms the bolometric picture. We refer also to the photocurrent study of Freitag et al., who showed that the bolometric contribution is dominant for the photocurrent generation in doped graphene under bias [4]. The responsivity of our device is $1.7 \cdot 10^{-6}$ A/W at 1065 nm and hence two orders of magnitude lower than of graphene [4].

Since the bolometric photocurrent is inversely proportional to the squared resistance [30], the lower responsivity is likely due to the one order of magnitude higher resistance of our device. We have demonstrated that nanocrystalline graphene can be used as a substitute to crystalline graphene for bolometric light detection.

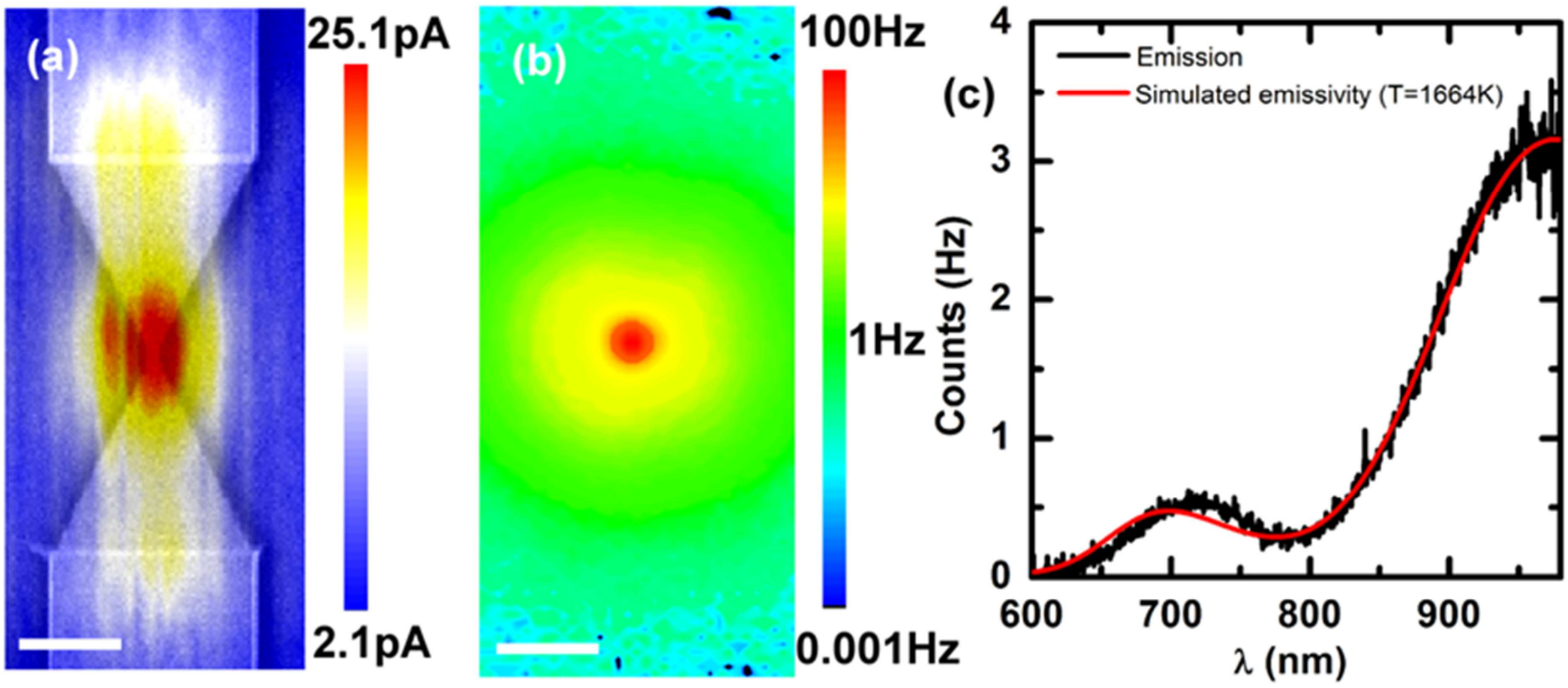

**Figure 6**: 1 nm thin NCG layer, patterned within two opposing NCG triangles connected by a 0.5 µm wide and 1 µm long NCG constriction. (a) Photocurrent map. $V_{SD}$ = -1V. Scalebar equals 5 µm. (b) Light emission image from a contact at $V_{SD}$ = 33V and $I_{SD}$ = 150 µA. Scale bar equals 5 µm. (c) Emission spectrum recorded at the center of (b).

We will now report the first study on light emission from nanocrystalline graphene and discuss the mechanism of light generation. Figure 6b shows a device with a NCG constriction under 33 V applied bias voltage and 150 µA current. The dimensions of the sample are identical to the one in Figure 6a. We observed a local light emission from the constriction, which we have identified previously as a region of high resistance (note that a logarithmic intensity scale has been used). As in the case of graphene we anticipate the source of light emission to be thermal radiation [5]. We have measured the spectrum of the emission which is shown in Figure 6c. The intensity overall increases towards higher wavelength however with obvious modulations. The data could be fitted nicely with a Planck curve and a structure dependent interference term (800nm-$SiO_2$/Si). Details can be found in the supporting information of [31]. We have extracted an electron temperature $T$ = 1664 K. If we assume that the electrical power is mainly dissipated at the NCG constriction we can estimate a local power density $p$ of 1000 kW/cm$^2$. These numbers are very similar to $T$ = 1600 K at $p$ = 520 kW/cm$^2$ reported for graphene [5]. We have therefore demonstrated that nanocrystalline graphene can be used as a substitute to crystalline graphene for generating thermal light emission on the nanoscale with comparable efficiency. Related to our previous work on narrow-band thermal light emission from microcavity-integrated graphene [8], we are currently replacing graphene with NCG to make the entire manufacturing compatible to wafer scale processing.

Finally we will discuss a piezoresistive effect that we have been able to measure on nanocrystalline graphene. Figure 7a shows a sample with a 1 nm thin double T-shaped NCG layer on 800nm-$SiO_2$/Si. The sample was loaded head-over into a 3-point bending fixture after electrical wiring (Fig. 7b). The resistance had been measured continuously while increasing the strain by vertical displacement of the inner post of the fixture. Data was acquired under increasing strain until sample failure. Figure 7c shows the relative resistance change ΔR/R as a function of the strain ε for two samples. The samples show a linear increase of ΔR/R until the breaking of the substrate at ε ≈ 0.1%. The linear slope corresponds to the gauge factor and is on the order of 20 for both samples. Whether the linearity continues towards larger strain values is subject to a future study and requires the transfer of NCG onto a flexible substrate. Nevertheless we can compare the data of NCG with measurements on CVD graphene as shown in figure 7d. We can see that the relative resistance change is larger in NCG than in graphene at a comparable strain. This obvious difference indicates that grain boundaries play an important role for the appearance of a piezoresistive effect in graphene. Indeed a recent calculation has shown that Klein tunneling should be absent for certain grain boundaries leading to strain-induced conductance modulation [32], which is piezoresistance. If the gauge values continue to be high also at high strain values, then the fabrication of transparent strain sensors based on NCG can be envisioned, which could offer new functionalities to the emerging field of transparent and flexible electronics.

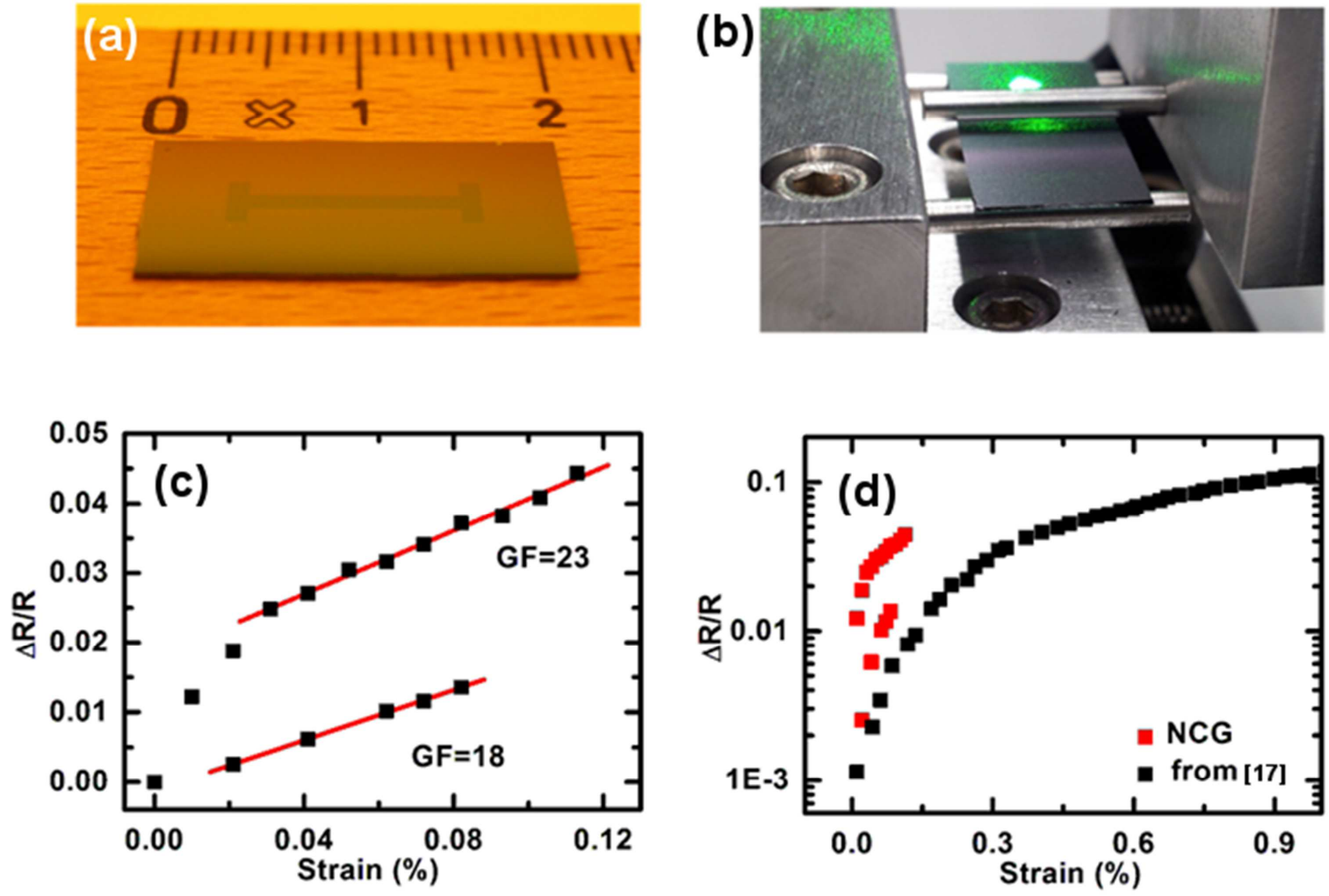

**Figure 7**: (a) Optical micrograph of a patterned 1 nm thin NCG layer on 800nm-$SiO_2$/Si suitable for piezoresistance measurements. (b) 3-point bending fixture with head over inserted sample. The movable piston and the adjustment laserspot are visible, the electrical contacts not. (c) Relative resistance change ΔR/R in NCG versus strain for two samples. The linear fits and corresponding gauge factors are indicated. (d) Comparison between ΔR/R of NCG with CVD-graphene from ref [19].

## 4. Conclusion/Summary

We demonstrated the wafer-scale synthesis of nanocrystalline graphene on dielectric surfaces by graphitization of a photoresist under high vacuum annealing, where the thickness, the sheet resistance and the transparency of the layer was tailored by the process condition. The layer is entirely formed by sp2-hybridized carbon as proven by XPS and Raman. The size of the graphitic domains is on the order of 2 nm, consistent with Raman and TEM measurements. Integrated into devices, the material showed photocurrent generation under illumination. The response to light could be traced to a bolometric origin, similar to experiments on doped crystalline graphene. Also light emission under electrical biasing was observed. The emission is due to heating of the layer, and the extracted electron temperature and power density is comparable to experiments reported from crystalline graphene. Furthermore a piezoresistive effect was observed that is significantly larger than in crystalline graphene and indicates the importance of grain boundaries for the appearance of piezoresistivity in graphene. Hence nanocrystalline graphene appears be an interesting material not only as an easy to fabricate alternative to crystalline graphene for nanoscale light detection and light generation but also towards the fabrication of transparent and flexible strain sensors.

**Acknowledgement**
RK acknowledges funding by the German Science Foundation (INST 163/354-1 FUGG), and RK and AR acknowledge funding by the VolkswagenStiftung. AF is supported by the Belgian fund for scientific research (FNRS).

**Supporting information**

Raman data of NCG layers synthesized at different temperatures (Figure S1), for different graphitization times (Figure S2), and on different Quartz substrates (Figure S3). NCG Raman D and G mode position, width and intensity ratio (Table T1). Scanning electron micrograph of laterally patterned NCG (Figure S4 and S5). EELS spectrum of NCG (Figure S6).

## Supporting information

Raman data of NCG layers synthesized at different temperatures (Figure S1), for different graphitization times (Figure S2), and on different Quartz substrates (Figure S3). NCG Raman D and G mode position, width and intensity ratio (Table T1). Scanning electron micrograph of laterally patterned NCG (Figure S4 and S5). EELS spectrum of NCG (Figure S6).

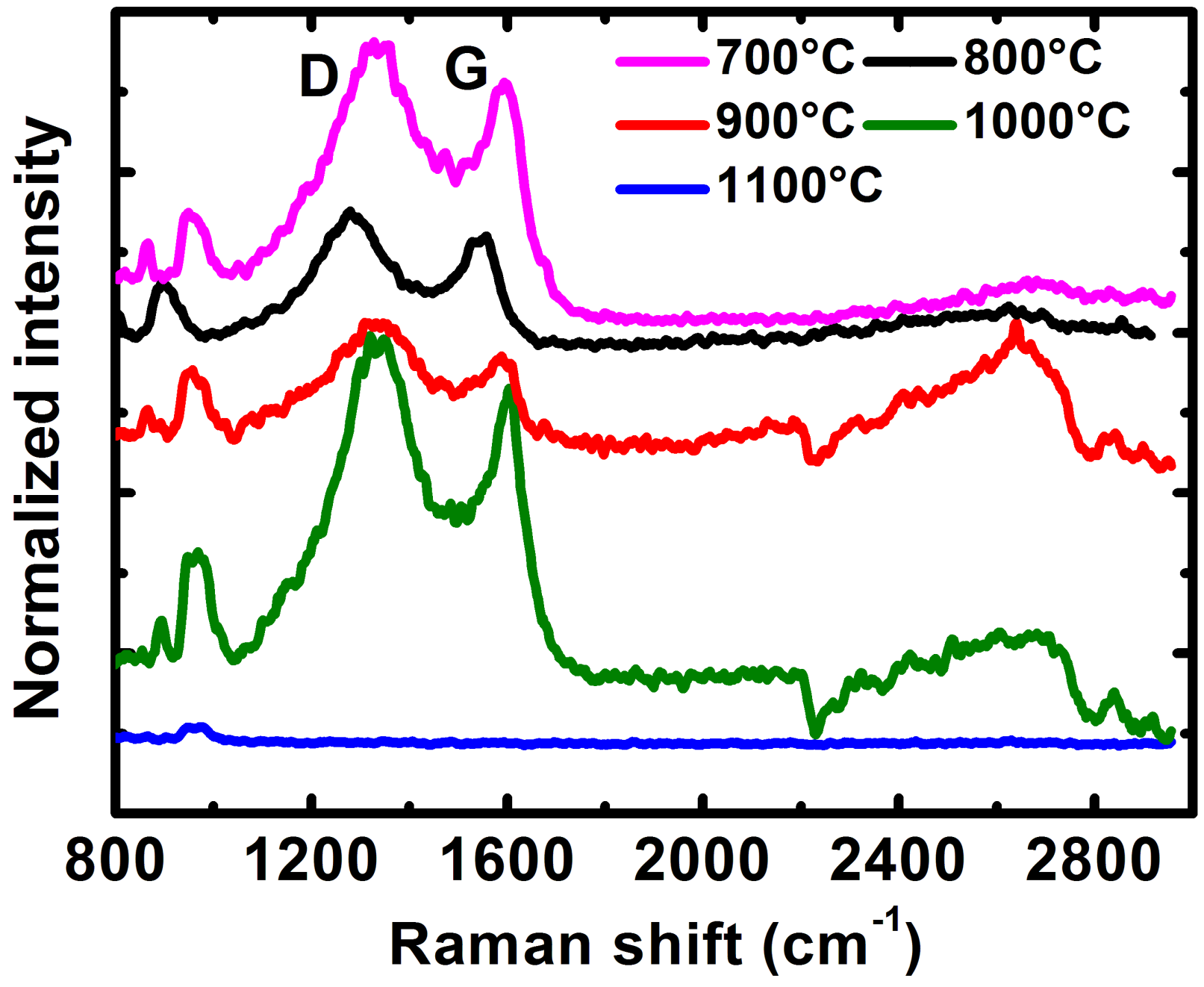

**Figure S1**: Raman data of 4nm NCG on 800nm-$SiO_2$/Si graphitized different temperatures for 10h.

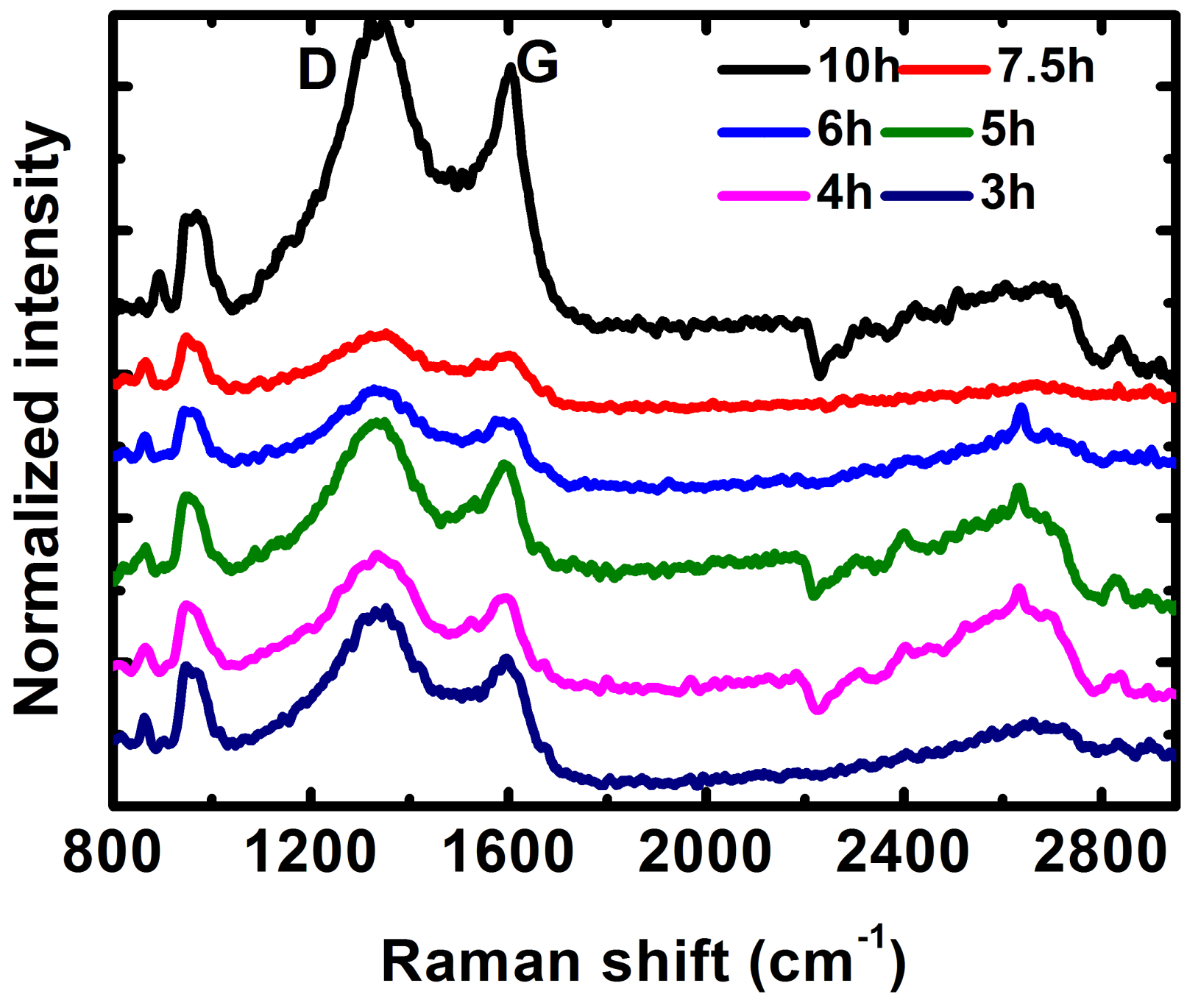

**Figure S2**: Raman data of 4nm NCG on 800nm-$SiO_2$/Si graphitized at 1000°C for different annealing times.

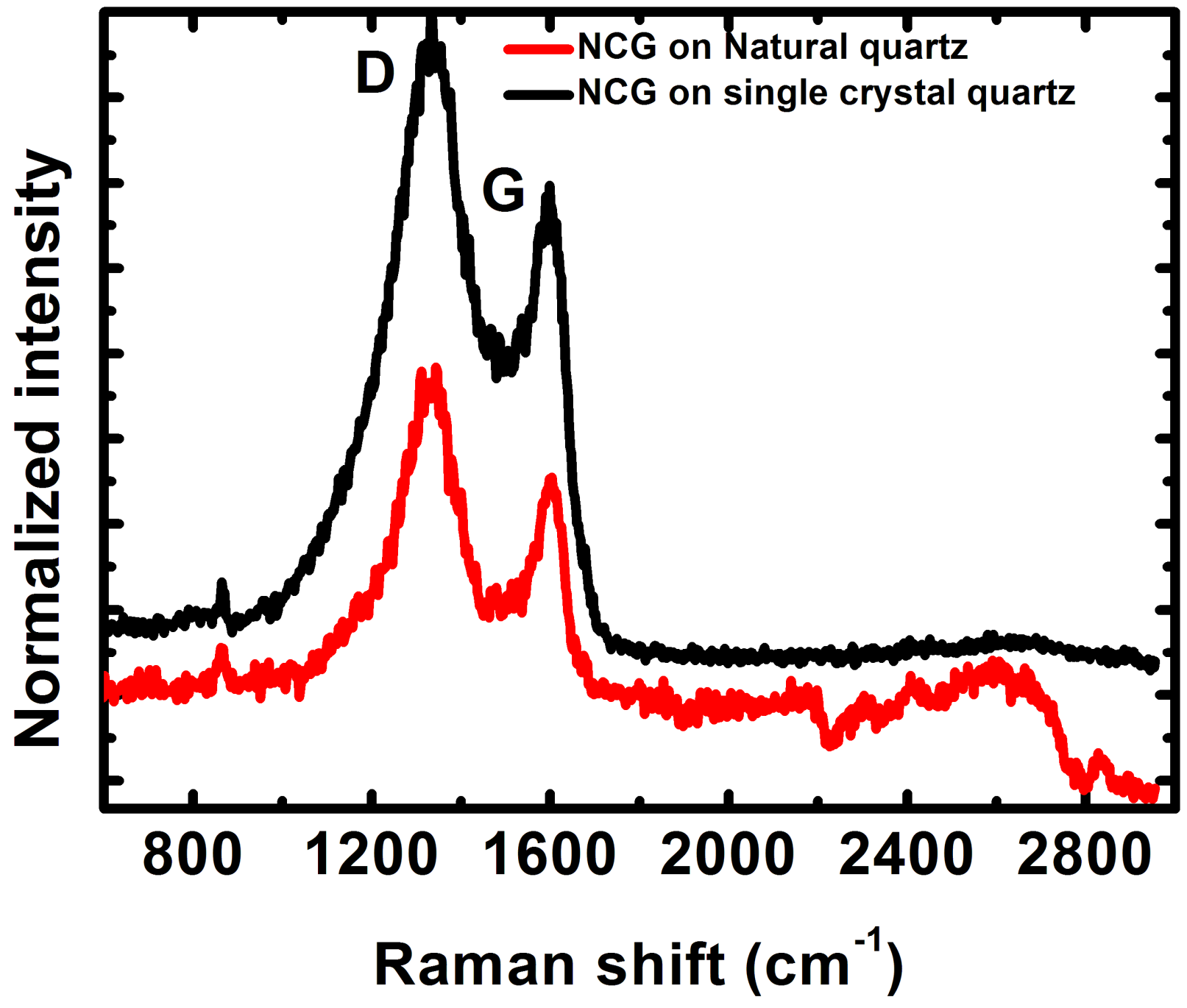

**Figure S3**: Raman data of 6nm NCG on quartz glass and on <0001> z-cut Quartz, graphitized at 1000°C@10h.

| NCG Thickness | Raman parameters | Substrates | | | | |
|---|---|---|---|---|---|---|
| | | Si/$SiO_2$ (100nm) | Si/$SiO_2$ (300nm) | Si/$SiO_2$ (800nm) | Single crystal Quartz | Natural Quartz |
| 6nm | D peak position ($cm^{-1}$) | 1342 | 1326 | 1332 | 1339 | 1333 |
| | G peak position ($cm^{-1}$) | 1605 | 1603 | 1601 | 1604 | 1598 |
| | $I_D$ (a.u) | 258 | 591 | 280 | 913 | 1873 |
| | $I_G$ (a.u) | 218 | 297 | 170 | 662 | 1304 |
| | $I_D/I_G$ | 1.2 | 2.0 | 1.6 | 1.4 | 1.4 |
| | FWHM $D_{peak}$ ($cm^{-1}$) | 220 | 164 | 284 | 211 | 120 |
| | FWHM $G_{peak}$ ($cm^{-1}$) | 99 | 74 | 80 | 96 | 107 |
| 4nm | D peak position ($cm^{-1}$) | 1342 | 1332 | 1329 | | |
| | G peak position ($cm^{-1}$) | 1600 | 1610 | 1600 | | |
| | $I_D$ (a.u) | 133 | 152 | 722 | | |
| | $I_G$ (a.u) | 109 | 119 | 563 | | |
| | $I_D/I_G$ | 1.2 | 1.3 | 1.3 | | |
| | FWHM $D_{peak}$ ($cm^{-1}$) | 222 | 148 | 258 | | |
| | FWHM $G_{peak}$ ($cm^{-1}$) | 101 | 59 | 105 | | |
| 2nm | D peak position ($cm^{-1}$) | 1342 | 1329 | 1326 | | |
| | G peak position ($cm^{-1}$) | 1604 | 1607 | 1603 | | |
| | $I_D$ (a.u) | 120 | 180 | 310 | | |
| | $I_G$ (a.u) | 104 | 136 | 208 | | |
| | $I_D/I_G$ | 1.2 | 1.3 | 1.5 | | |
| | FWHM $D_{peak}$ ($cm^{-1}$) | 207 | 154 | 249 | | |
| | FWHM $G_{peak}$ ($cm^{-1}$) | 94 | 67 | 119 | | |
| 1nm | D peak position ($cm^{-1}$) | 1341 | 1329 | 1330 | | |
| | G peak position ($cm^{-1}$) | 1605 | 1610 | 1599 | | |
| | $I_D$ (a.u) | 82 | 217 | 162 | | |
| | $I_G$ (a.u) | 69 | 175 | 120 | | |
| | $I_D/I_G$ | 1.2 | 1.2 | 1.4 | | |
| | FWHM $D_{peak}$ ($cm^{-1}$) | 175 | 200 | 198 | | |
| | FWHM $G_{peak}$ ($cm^{-1}$) | 92 | 76 | 80 | | |

**Table T1:** Raman data of samples synthesized at 1000°C on various substrates and NCG thicknesses.

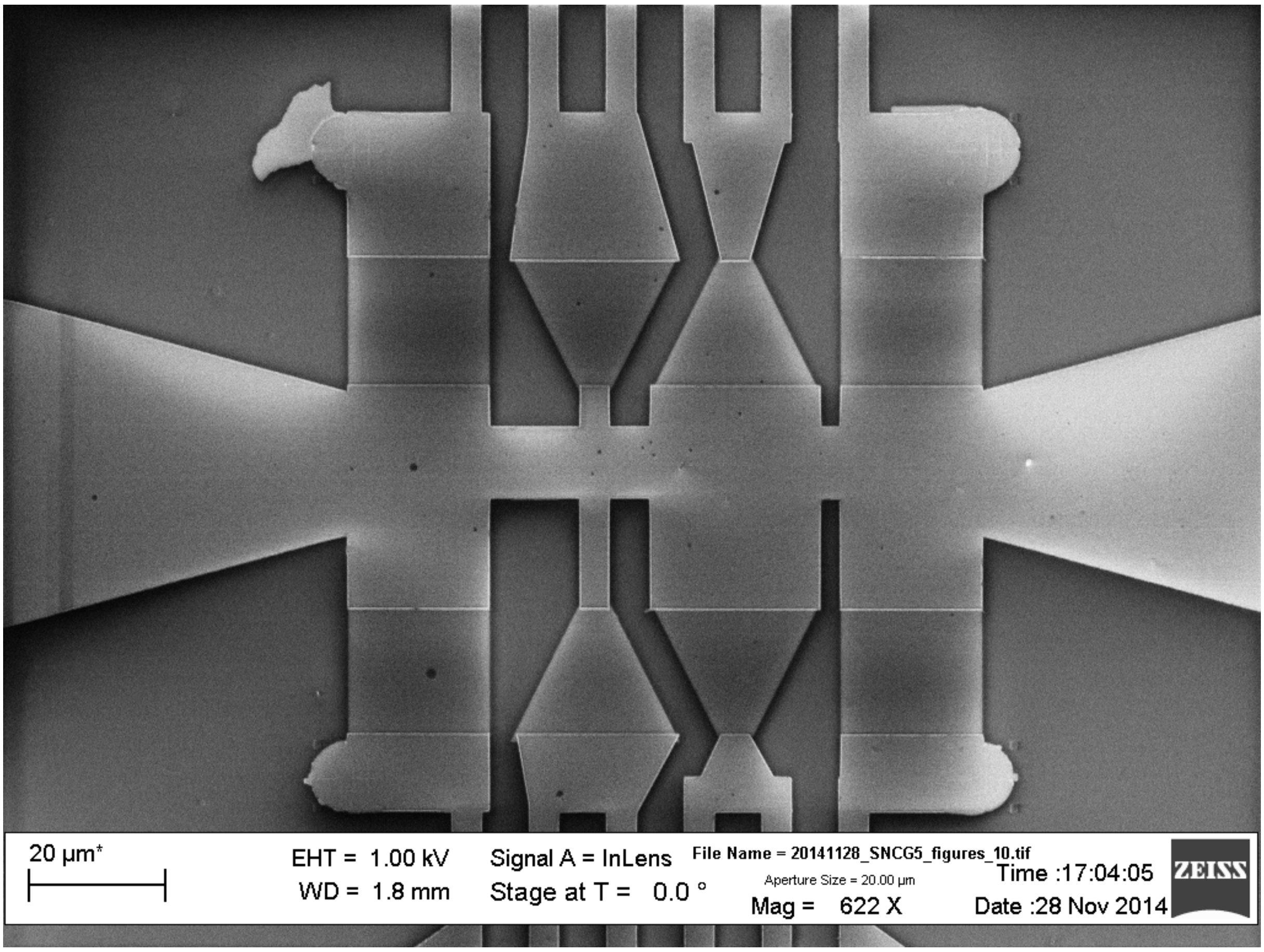

**Figure S4**: Scanning electron micrograph of 4 nm thin, square-shaped and triangle-shaped patterned NCG.

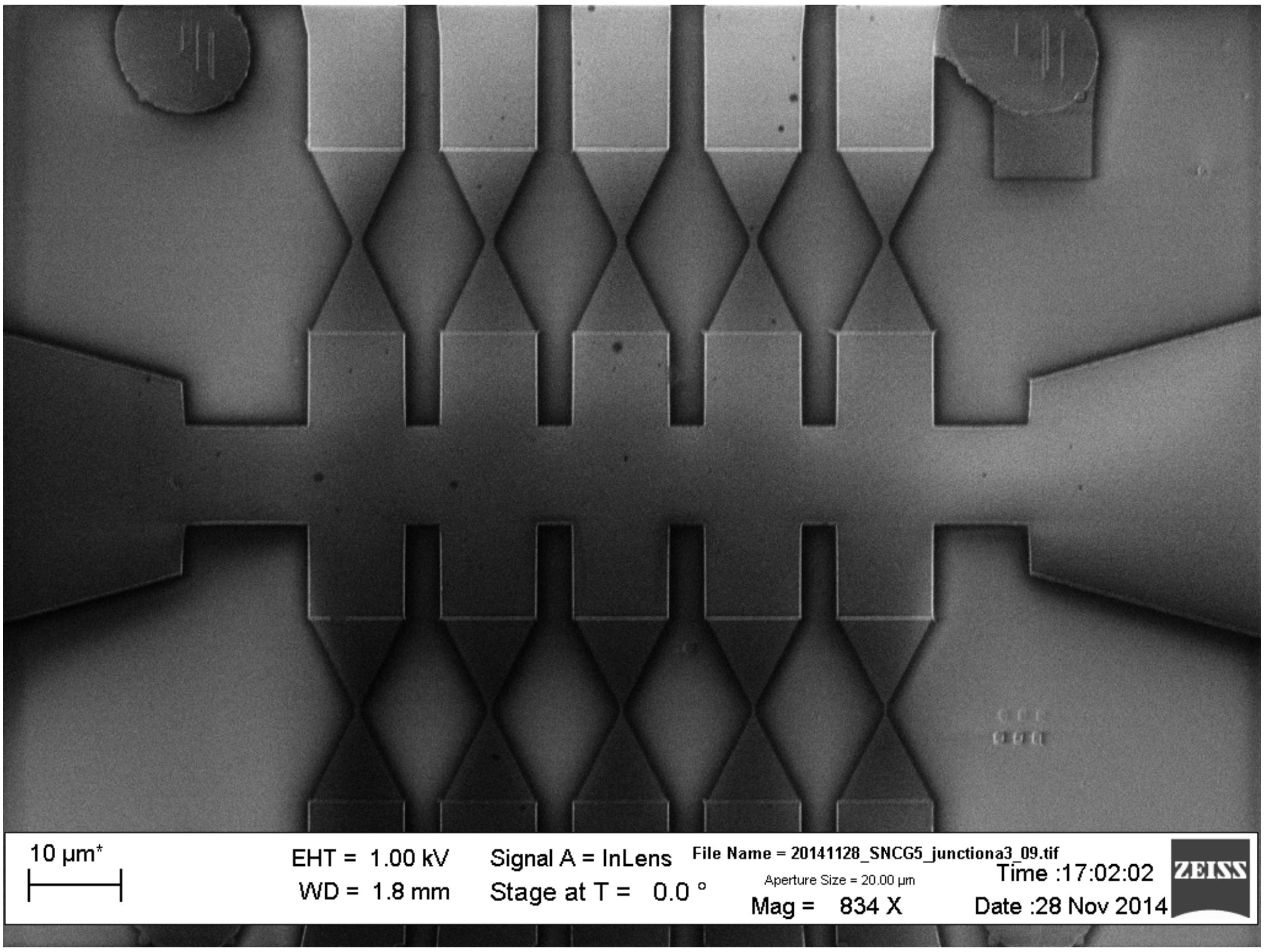

**Figure S5**: Scanning electron micrograph of 1 nm thin NCG layer, patterned within two opposing NCG triangles with 20 µm x 20 µm overall dimensions connected by a 0.5 µm wide and 1 µm long NCG constrictions.

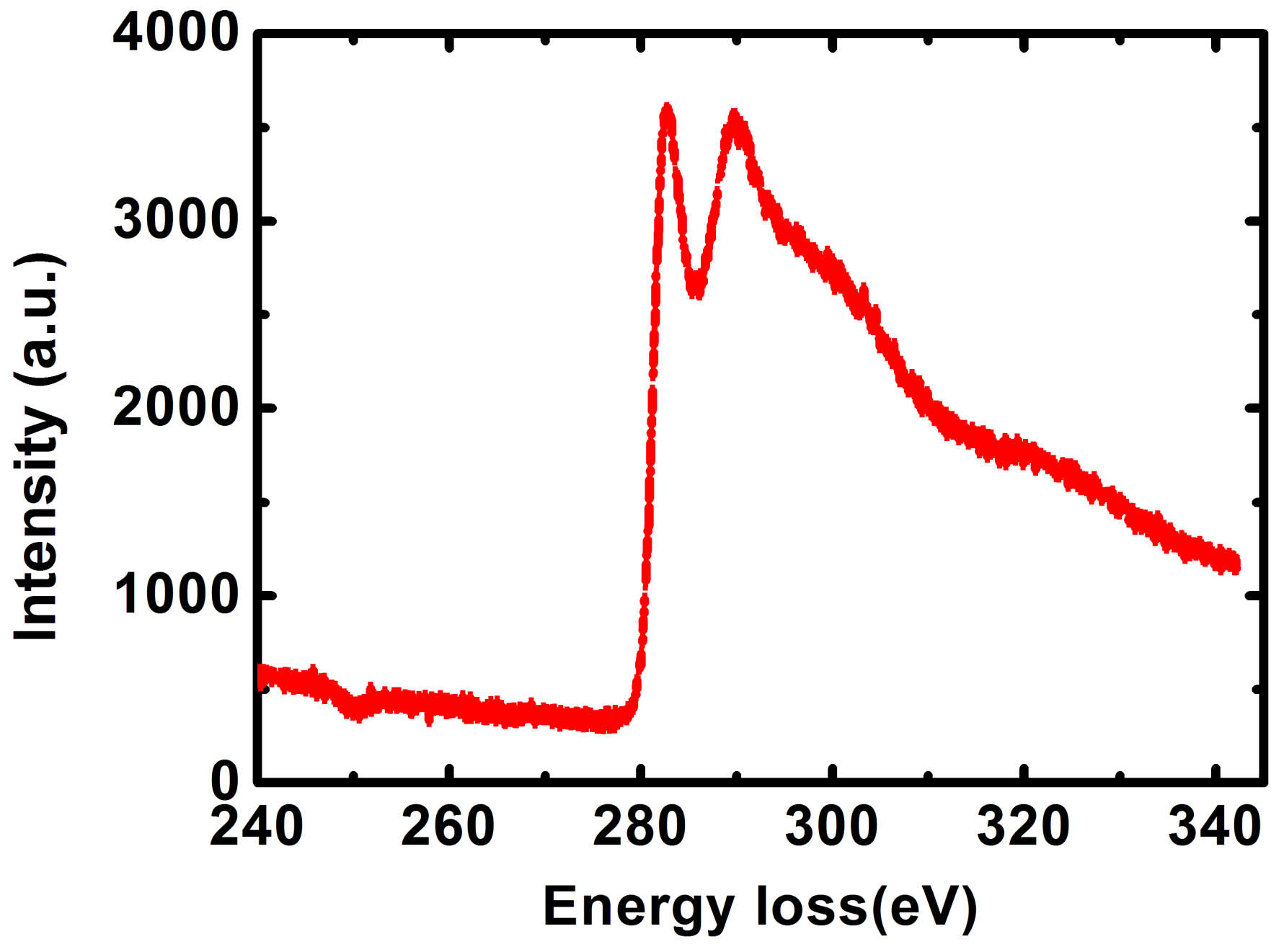

**Figure S6**: EELS spectrum of a suspended 2 nm thick NCG layer.